\documentclass[final,authoryear,5p,twocolumn]{elsarticle}

\usepackage{graphicx}

\usepackage{amssymb}

\def\astrobj#1{#1}

\journal{New Astronomy}

\begin{document}

\begin{frontmatter}



\title{A photometric and spectroscopic study of \astrobj{WW And} - an Algol-type, 
long period binary system with an accretion disc}

\author[1]{Michal Siwak\tnoteref{label1}}
\tnotetext[label1]{e-mail: siwak@oa.uj.edu.pl}
\author[1,2]{Stanislaw Zola}
\author[1]{Tomasz Szymanski}
\author[1]{Maria Kurpinska-Winiarska}
\author[1,2]{Maciej Winiarski}
\author[1]{Dorota Koziel-Wierzbowska}
\author[1]{Waclaw Waniak}
\author[3]{Michal Drahus}
\address[1]{Astronomical Observatory, Jagiellonian University, 
ul.\ Orla 171, 30-244 Krak{\'o}w, Poland,}
\address[2]{Mount Suhora Astronomical Observatory, Cracov Pedagogical University, 
ul.\ Pochor\c{a}{\.z}ych 2, 30-084 Krak{\'o}w, Poland}
\address[3]{Department of Earth and Space Sciences, University 
of California at Los Angeles, 595 Charles E. Young Dr. E.,
3711 Geology, Los Angeles, CA 90095-1567}

\begin{abstract}
We have analyzed the available spectra of \astrobj{\astrobj{WW And}} 
and for the first time obtained a reasonably well defined radial 
velocity curve of the primary star.
Combined with the available radial velocity curve of the secondary
component, these data led to the first determination of the spectroscopic 
mass ratio of the system at q$_{spec} = 0.16\pm 0.03$. 
We also determined the radius of the accretion disc from analysis of the 
double-peaked H$_\alpha$ emission lines.
Our new, high-precision, Johnson $VRI$ and the previously-available 
Str{\"o}mgren 
$vby$ light curves were modelled with stellar and accretion disc models. 
A consistent model for \astrobj{WW And} -- a semidetached system harbouring 
an accretion disc which is optically thick in its inner region, but optically thin
in the outer parts -- agrees well with both spectroscopic and photometric data.  
\end{abstract}


\end{frontmatter}



\section{Introduction}
\label{intro}

\astrobj{WW And} ($V=10.9~mag$, $B-V=0.49\pm0.09$, A5+F3p) is a long period 
($P_{orb} \approx 23.3~d$), Algol-type binary system. 
The first spectroscopic data obtained in the blue part
of the spectrum at orbital phases close to the primary minimum 
revealed red emission components in hydrogen 
and the Ca~II~K absorption lines \citep{wyse,struve}, and suggested 
the existence of H$_{\alpha}$ and H$_{\beta}$ emission.
As the source of the emission they proposed circumstellar matter 
lost by the Roche lobe filling secondary star and accreted by the
primary component through the first Lagrangian point. 
Due to the considerable contamination of these lines by the circumstellar 
matter, the radial velocity curve of the primary star remained undefined \citep{struve}.

Direct evidence for an accretion disc surrounding the primary, 
mass-gaining, component came with the discovery of double-peaked 
H$_{\alpha}$ emission line \citep{Olson93}. 
A more detailed analysis of the line profile reveals that changes 
of the equivalent width (EW) of its violet and red emission components 
with orbital phase can be explained by enhanced emission from 
the stream-impact region localized in the accretion disc \citep{olson95}. 
An additional argument for the presence of a large accretion disc
follows from analysis of EW changes of 
the O~I~7774 absorption line over the orbital period \citep{etzel95}.

The first photometric light curves of \astrobj{WW And} in $uvby$-Str{\"o}mgren 
and $I$-Kron filters were obtained by \citet{Olson93}. 
They made an attempt to model the light curves with the
Wilson-Devinney (WD) light curve synthesis code \citep{wilson79}, 
assuming a semidetached configuration. However, a significantly better fit 
was obtained for a detached geometry, with the secondary component 
filling about $93\%$ of its Roche lobe. 
The above-mentioned arguments indicating the presence
of an accretion disc around the primary component  
led the authors to conclude that mass loss from the secondary star 
occurs through the stellar wind rather than Roche lobe overflow.
\citet{zola97b} modelled the \astrobj{WW And} $vbyI$ light curves assuming a semidetached configuration and using
a modified WD code accounting for effects 
introduced by an accretion disc around the primary component. 
A reasonable fit was obtained but the disc radius was smaller than 0.1 
of the orbital separation and the solution was not unique -- the fit was 
almost equally good for a range of photometric mass ratio values.

In order to get a definitive answer about the configuration 
of this system and obtain more accurate properties of the accretion disc 
from light curve modelling, we decided to obtain new, high-quality 
photometric data especially at phases around the secondary minimum. 
To minimize the scatter due to accretion effects observed in many other mass 
transferring Algol-type binary systems, a large effort was made to collect 
a complete light curve in Johnson $BVRI$ filters in as short a time as possible.
We describe the new photometric data in Section~\ref{obs} and  
compare the new Johnson $V$ and the older Str{\"o}mgren $y$ light 
curves in order to check for possible changes of the primary 
minimum depth.
In Section~\ref{spec} we present the results obtained from analysis of 
the spectroscopic data -- the spectroscopic mass 
ratio (determined for the first time) and the size of the accretion disc radius. 
Models obtained from $vbyVRI$ light curve modelling are presented 
in Section~\ref{lcm}.
We summarize the results in Section~\ref{discus}.

\section{Observations and data reduction}
\label{obs}

We observed \astrobj{WW And} with the 50~cm Carl~Zeiss telescope 
at the {\it Fort Skala} Astronomical Observatory 
of the Jagiellonian University.
This telescope was equipped with a Photometrics S300 CCD camera 
with the SITe SI003B, 1024x1024-pixel chip and a set of broadband 
Johnson-Bessell $UBVRI$ filters, manufactured by Custom Scientific.
With the 6.7~m focal length of this telescope, the field of view 
was 12'x12'.

The data were collected during 86 nights over 4 consecutive seasons 
in $BVRI$ filters. The first observations were obtained 
on 4/5th January, 2002, while the latest ones on 21/22 October, 2005.
They were reduced for {\it bias}, {\it dark} and {\it flatfield} 
in the usual way using our scripts working under the {\it ESO-MIDAS} software 
environment.
The aperture photometry was made making use of {\it DAOPHOT~II} package \citep{stet}.
GSC 3638-01674 (having almost the same colour index as the variable 
star) served as the comparison, while GSC 3638-01421 and GSC 3638-01742 
as the check stars.
No night-to-night variations larger than 0.01~mag ($\sigma$)
were noticed in the {\it comparison-check} star differential photometry 
in all filters. 
Due to the small angular separation between the variable, comparison 
and check stars, the corrections for differential 
atmospheric extinction $k$ turned out to be negligible over the entire 
range of observed airmasses.
The data obtained in the $B$ filter were also corrected for colour 
extinction $\beta$ using the mean coefficient obtained for 
the {\it Fort Skala} Observatory by Winiarski (private communication).
As a result, we obtained a good quality light curves with 
a formal scatter of the individual points of a few milimagnitudes.
As the secondary minimum was observed several times it is much better 
covered than in the {\it uvbyI} light curves obtained 
by \citet{Olson93}.
The new data show a scatter whose amplitude grows 
toward shorter wavelengths, reaching almost 0.02 mag in the $B$ filter. 
An analogous trend is also visible in the {\it uvbyI} light 
curves \citep{Olson93}.
\newline 
The data were left in the instrumental system and the phases
were calculated using the most recent linear elements 
provided by J.M. Kreiner \citep{kreiner04}:

\begin{equation}
HJD = 2,453,665.502 (23) \ + \ E \times 23.28525 (3), 
\end{equation}
where $E$ is an epoch number.  

According to visual observations made at the beginning of the
20th century and presented by \citet{leiner}, the primary minimum of
\astrobj{WW And}, lasting about 0.07~$P_{orb}$, was total and nearly 1.1~mag deep.
As the Str{\"o}mgren $y$-band data clearly shows a shorter (0.05~$P_{orb}$), partial, 0.6~mag 
deep primary eclipse, \citet{Olson93} 
proposed that perhaps \astrobj{WW And} continuously changes its orbital inclination. 
However, \citet{meyer} re-analyzed Leiner's data and gave a strong 
proof that the depth of minimum given by Leiner was erroneous and most probably the suggested change of inclination 
is not taking place.
\newline
In order to check the above hypothesis by means of more uniform data set, 
we compared the Olson's $y$ and our $V$ light curves, obtained 15--20 years 
later. 
We found that primary minimum in our new $V$ data is only 0.02~mag deeper than that in the $y$ filter. 
Such a small difference can be fully explained by combination of
small changes in the circumbinary environment and the lack
of transformation between the $V$ and $y$ bands. 

\begin{figure*}
\centerline{%
\begin{tabular}{c@{\hspace{1pt}}c@{\hspace{1pt}}c@{\hspace{1pt}}c}
\includegraphics[width=1.22in, angle=-90]{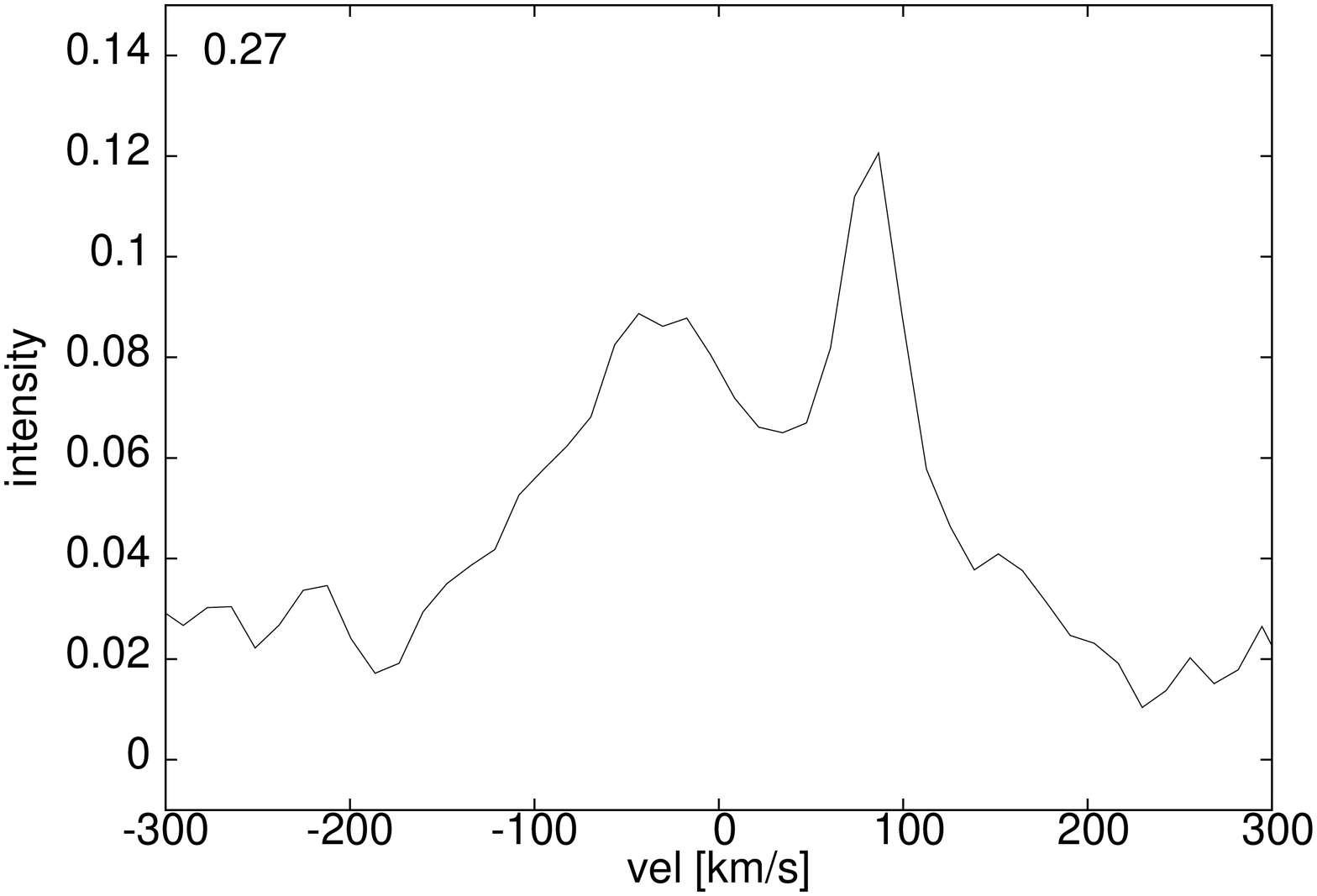} &
\includegraphics[width=1.22in, angle=-90]{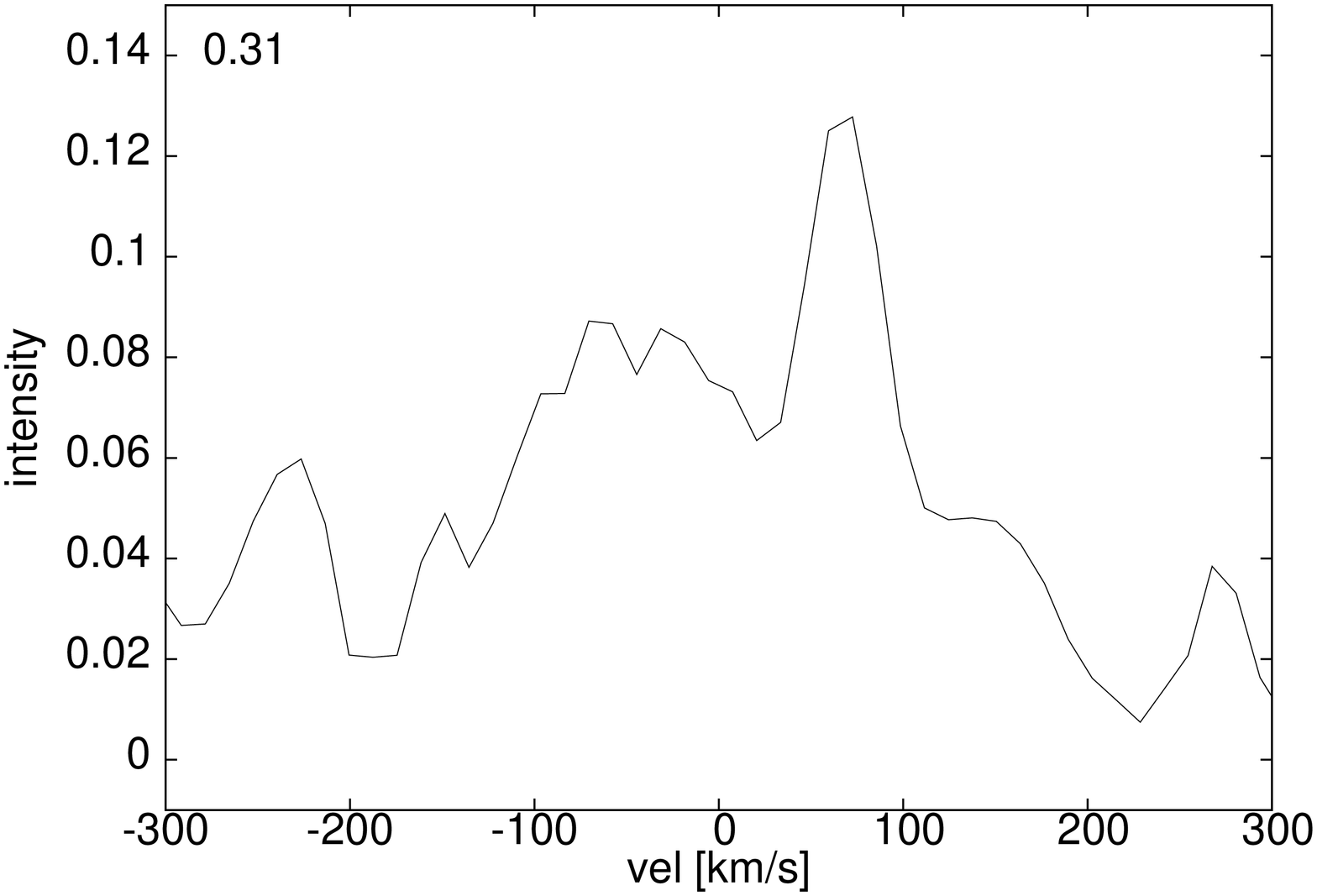} &
\includegraphics[width=1.22in, angle=-90]{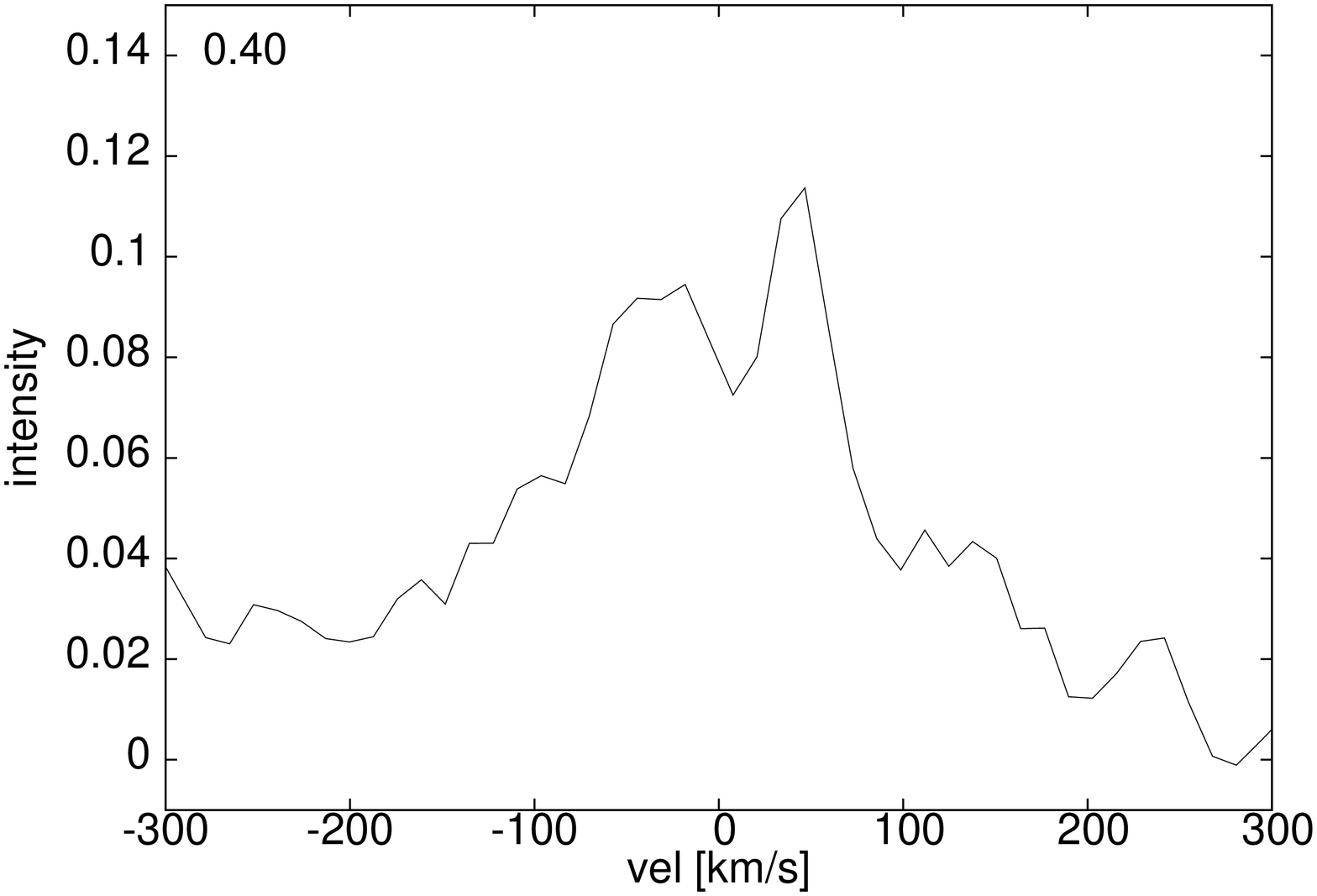} &
\includegraphics[width=1.22in, angle=-90]{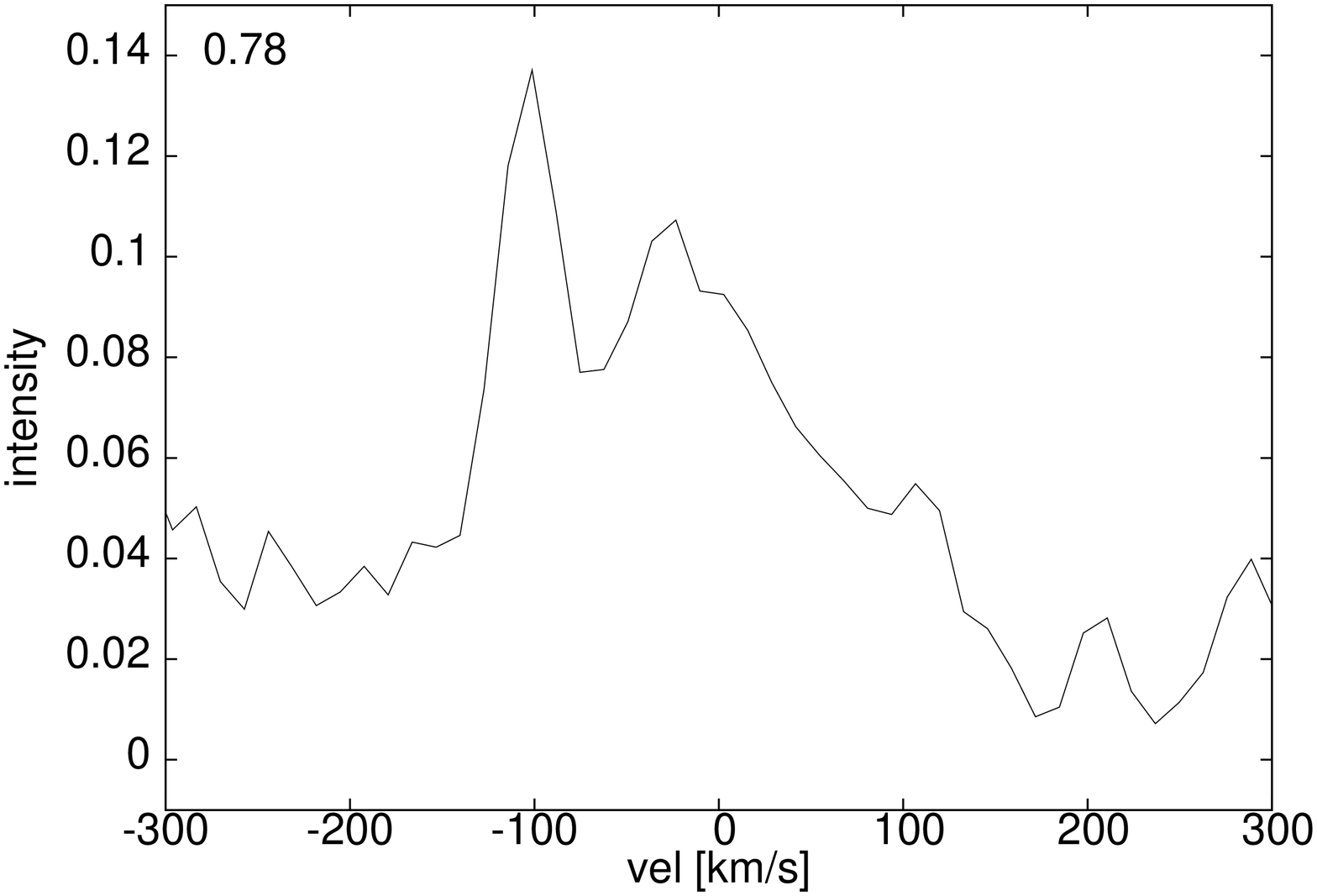}
\end{tabular}}
\caption {Broadening Functions obtained from spectra obtained 
in Mg II 4481~\AA~region in different orbital phases, indicated 
in plots.}
\label{Fig.1}
\end{figure*}
\begin{table}
\caption{Radial velocities of both components (RV$_1$ - primary, RV$_2$ - secondary) 
obtained by means of the Broadening Function method from spectra obtained 
by \citet{Olson93} in the Mg~II~4481\AA~region. 
The accuracy of individual data points is no worse than 5~km/s (see Sec.~\ref{bf}).}
\vskip 5pt
\centering{
\begin{tabular}{r c r r}
 \hline\hline
HJD-2,448,000 &  phase   & RV$_1$ [km/s] & RV$_2$ [km/s] \\ \hline
 456.83665    &  0.3107  & -37.4      & 68.7         \\
 456.86443    &  0.3119  & -31.0      & 64.9         \\
 458.88888    &  0.3989  & -28.9      & 40.6         \\
 851.75641    &  0.2708  & -25.9      & 81.9         \\
 851.79452    &  0.2725  & -24.5      & 80.2         \\
 886.95239    &  0.7823  &  -2.3      & -102.5       \\
1180.89606    &  0.4060  & -21.7      & 46.3         \\ \hline \hline
\end{tabular}
\label{Tab.1}
}
\end{table}

\section{Analysis of spectroscopic data}
\label{spec}

\subsection{Radial velocity determination}
\label{bf}

In Figure~\ref{Fig.1} we present four Broadening Functions (BF) calculated from 
the medium resolution (0.2\AA/pix) spectra obtained by \citet{Olson93} 
in the Mg~II~4481\AA~region. 
These reveal the sharp, well defined profile 
of the secondary component (as expected in the case of synchronous 
rotation and the long orbital period) and the broad profile 
of the primary star.
Due to the poor velocity sampling of the BF profiles ($13 \ km \ s^{-1}$) 
they cannot be studied in greater details; this would 
require high-resolution spectra.
 
We attempted to fit the broad, triangular profiles of the primary 
with rotationally broadened synthetic profiles, but failed.
A similar problem was encountered by \citet{prib08}
for the primary component of the close binary \astrobj{AW UMa}. 
Therefore, we tentatively propose that this shape of the profile is caused 
by the fast rotating, optically thick inner part of an accretion 
disc, rather than by close-to-critical rotation of 
the primary component, as proposed by \citet{Olson93}.
In Figure~\ref{Fig.1} we also notice that the BF profile of the primary 
component obtained at phase 0.782 is asymmetric, i.e. its maximum 
intensity is considerably shifted to smaller velocities. 
This may be explained by combination of effects 
introduced by a non-uniform flux distribution of the inner parts of 
the accretion disc, and/or by radiation from a hot spot produced by a stream-disc 
interaction, which is expected to be best visible at the second quadrature.

\begin{figure}
\includegraphics[width=60mm,angle=-90]{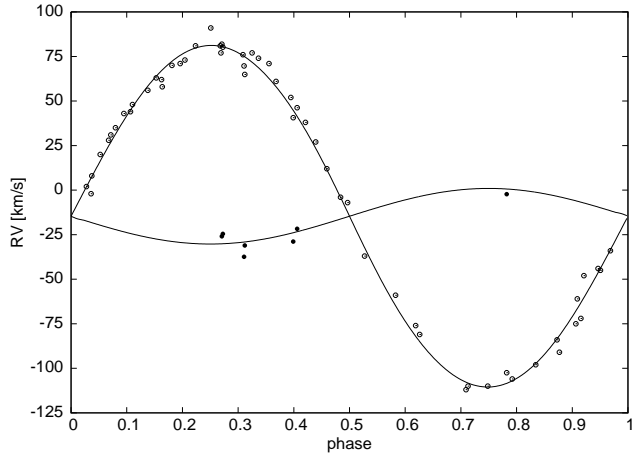}
\caption{The radial velocity curves of \astrobj{WW And} and the fit obtained with 
the WD code assuming a stellar model.
}
\label{Fig.2}
\end{figure}

From measurements of the {\it light centroids} of the BF profiles we obtained 
seven new radial velocity points for each component (Table~\ref{Tab.1}). 
Despite their uncertainty ($\approx 5 km/s$) related to limited 
instrument stability (Olson, private communication), they provide the first 
reasonably well defined radial velocity curve of the primary component 
(Fig.~\ref{Fig.2}). 
We appended our measurements to the data of the secondary component obtained 
from the iron lines in the H$_{\alpha}$ region \citep{Olson93} and 
as the result we obtained the spectroscopic value of the 
mass ratio, $q_{spec}=M_2/M_1=0.16\pm0.03$. This value is small enough 
to avoid direct stream impact on the primary component photosphere; 
instead the stream is expected to encircle the primary star and to form 
an accretion disc \citep{lubow75}.

\subsection{Disc radius determination}
\label{hmod}

\begin{figure}
\includegraphics[width=60mm,angle=-90]{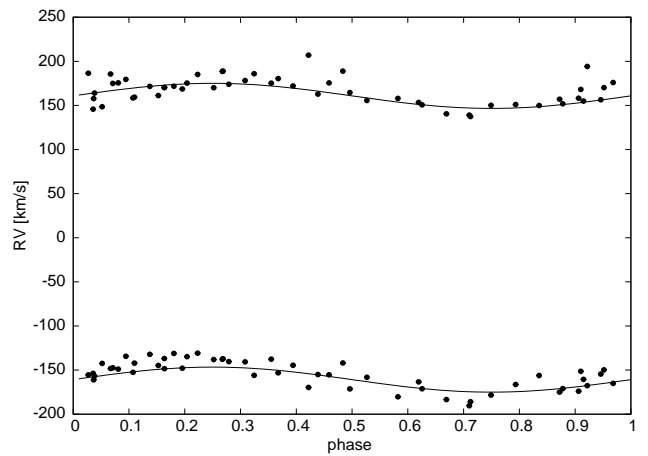}
\caption{Measured radial velocities of the H$_\alpha$ blue and red
peaks (points). Also shown the model for assumed keplerian rotation
of a circular disc (continuous lines).
}
\label{FigDiskPeaks}
\end{figure}

\begin{figure*}
\centerline{%
\begin{tabular}{c@{\hspace{1pt}}c@{\hspace{1pt}}c@{\hspace{1pt}}c}
\includegraphics[width=1.22in, angle=270]{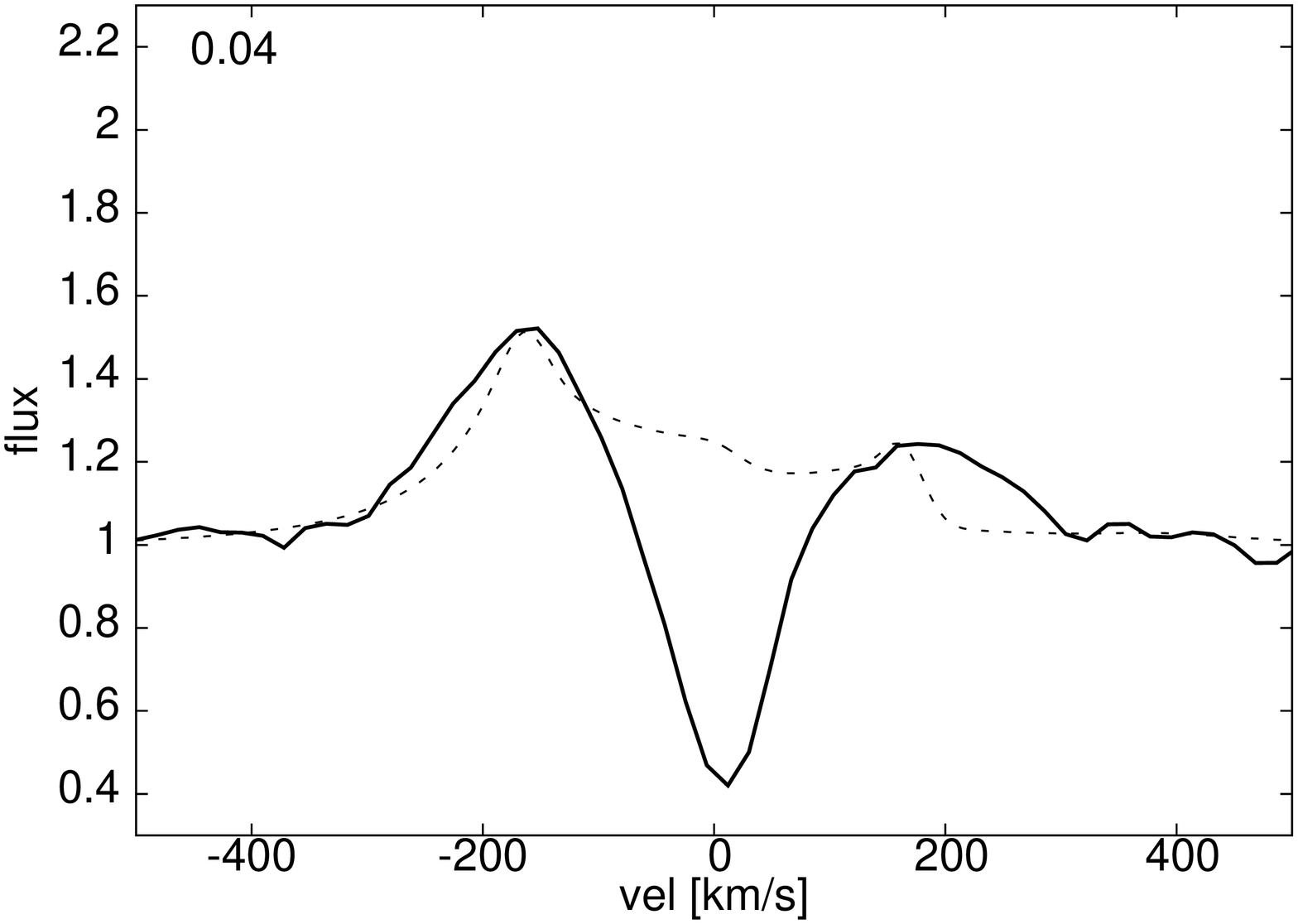} &
\includegraphics[width=1.22in, angle=270]{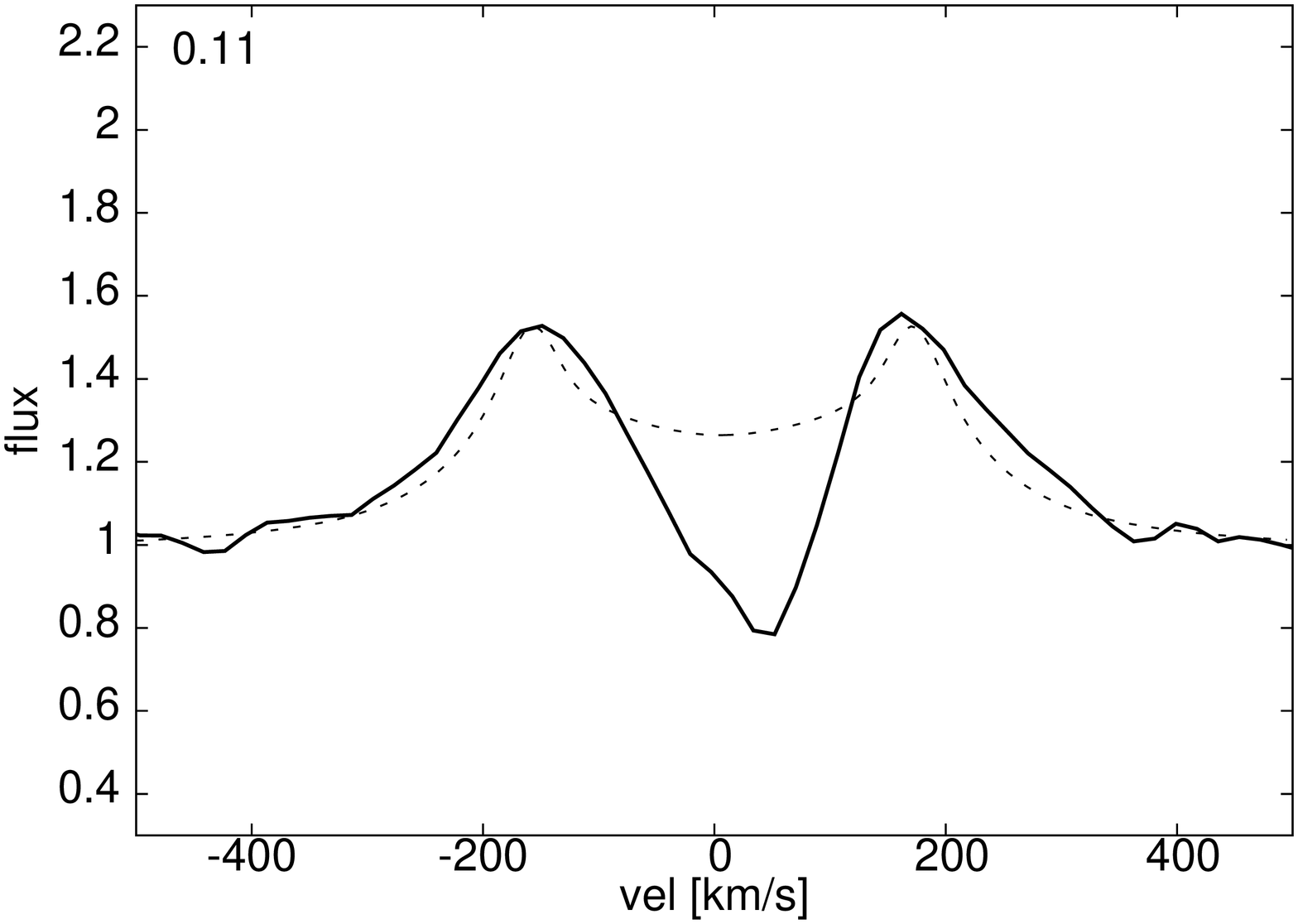} &
\includegraphics[width=1.22in, angle=270]{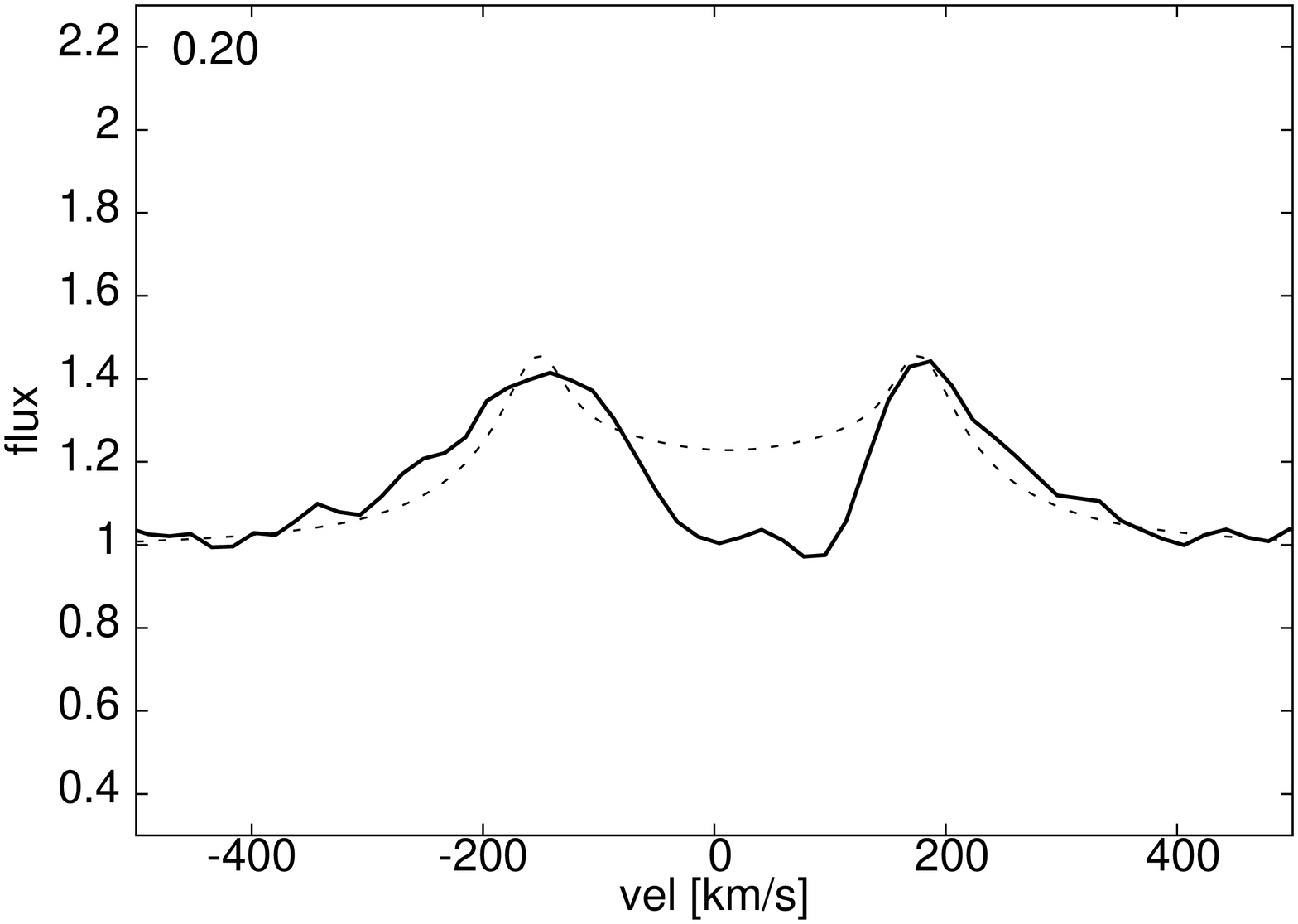} &
\includegraphics[width=1.22in, angle=270]{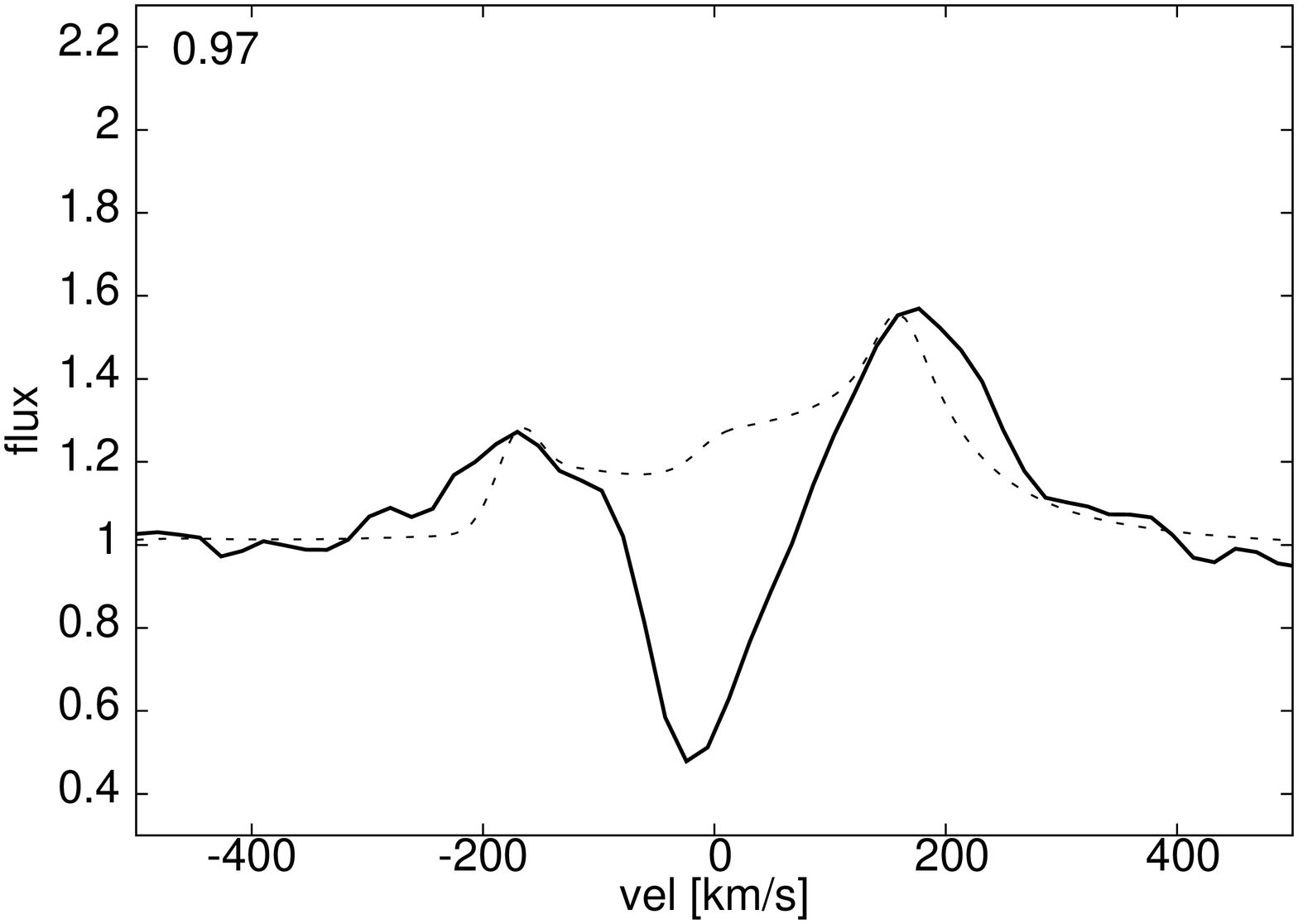}
\end{tabular}}
\caption {Comparison of the H$_\alpha$ line theoretical profiles (dotted line) 
with observations (continuous line). The phases are indicated in plots.} 
\label{FigHalphaProfiles}
\end{figure*}

The size of an accretion disc can be determined from separation 
of the blue and red peaks of the H$_\alpha$ emission line \citep{Smak1981}.
Therefore the H$_\alpha$ spectra obtained by \citet{Olson93} have been 
transformed into velocity space, accounting for the mean system velocity. 
While measuring the velocities of both peaks (Fig.~\ref{FigDiskPeaks}) 
it turned out that they are also contaminated by an absorption feature. 
We measured the radial velocities of this absorption line and found that they follow the RV 
curve of the secondary component. Based on this result we identify this
as the H$_\alpha$ line originating in the secondary. 
The derived size of the accretion disc as a function of inclination 
is given in Table~\ref{TabDiskRad}, while the synthetic profiles 
at four orbital phases are shown in Fig.~\ref{FigHalphaProfiles}. 

\begin{table}
\caption{Circular disc radius as a function of inclination (in units of component separation). 
}
\vskip 5pt
\centering{
\begin{tabular}{cr}
\hline\hline
inclination  &   disc radius \\
  \hline
 90    &  0.443     \\
 88    &  0.442     \\
 86    &  0.440     \\
 84    &  0.438     \\
 82    &  0.434     \\
 80    &  0.430     \\
 78    &  0.423     \\ 
 76    &  0.417     \\ 
 74    &  0.409     \\ 
\hline\hline
\end{tabular}
\label{TabDiskRad}
}
\end{table}

\section{Light curve modelling}
\label{lcm}

Due to significant scatter and seasonal effects visible in the $B$ data, 
this light curve was discarded from the light curve modelling.
From the original {\it VRI} data we calculated 140, 147, and 149 mean points, 
respectively and  145 mean points were calculated for each of the Str{\"o}mgren 
{\it vby} filters. 
The $vbyVRI$ data, converted into flux units and normalized to about unity 
at the phase 0.25, were then simultaneously modelled by means of the stellar 
and the accretion disc models.

\begin{table}
\centering
\caption {The best-fit values obtained from light curve modelling within the stellar and the accretion disc models. 
The errors are given in parentheses. The accretion disc temperature $T_d$ did not converge to any specific value.
\newline
Parameter: $^{*}$ - fixed, $^{**}$ - computed }
\vskip 5pt
\begin{tabular}{@{}lll@{}}
\hline\hline
parameter           & stellar model& disc model   \\ \hline
$i[^{\circ}]$       & 79.90 (1)    & 76.92 (11)   \\
$T_{eff}^{prim}[K]$ & 8150$^*$     & 8150$^*$     \\
$T_{eff}^{sec}[K]$  & 4135 (1)     & 4500$^*$     \\
$\Omega_1$          & 28.08 (5)    & 16.44 (67)   \\
$\Omega_2$          & 2.2330 (6)   & 2.1300$^{**}$\\ 
$q_{spec}$          & 0.16$^*$     & 0.16$^*$     \\ \hline
$L_{1v}$            & 7.420 (14)   & 2.04 (1.65)  \\
$L_{1b}$            & 6.531 (13)   & 3.11 (1.45)  \\
$L_{1y}$            & 5.343 (11)   & 3.63 (1.21)  \\
$L_{1V}$            & 5.259 (11)   & 3.46 (0.81)  \\
$L_{1R}$            & 4.098 (10)   & 3.67 (0.61)  \\
$L_{1I}$            & 2.742 (07)   & 3.18 (0.41)  \\ \hline
$L_{2v}$            & 5.187$^{**}$ & 1.11$^{**}$  \\
$L_{2b}$            & 5.888$^{**}$ & 2.13$^{**}$  \\
$L_{2y}$            & 6.965$^{**}$ & 3.39$^{**}$  \\
$L_{2V}$            & 6.943$^{**}$ & 3.26$^{**}$  \\
$L_{2R}$            & 7.978$^{**}$ & 4.83$^{**}$  \\
$L_{2I}$            & 9.176$^{**}$ & 6.66$^{**}$  \\ \hline
$l_{dv}$            & ---          & 0.771 (163)  \\
$l_{db}$            & ---          & 0.615 (160)  \\
$l_{dy}$            & ---          & 0.482 (157)  \\
$l_{dV}$            & ---          & 0.495 (105)  \\
$l_{dR}$            & ---          & 0.357 (92)   \\
$l_{dI}$            & ---          & 0.241 (79)   \\ \hline
$r_d$               & ---          & 0.084 (8)    \\
$T_d$               & ---          & not conv.    \\
$\gamma$            & ---          & 1.0$^*$      \\
$\beta$             & ---          & 6.2 (3.3)    \\ \hline  
$\chi^2_{red,w}$    &  5.153       & 3.023        \\ \hline\hline
\end{tabular}
\label{Tab.3}
\end{table}

\subsection{Stellar model}
\label{wd}

As the first step we modelled the $vbyVRI$ light curves using the 
1996 version of the WD code extended with the Monte Carlo method 
as a global minimum search procedure \citep{zola97, zola10}.
This was aimed at checking the system configuration, which in our code
is not assumed {\it a priori} but is obtained as a result. 
We also wanted to check how good the fit could be for the new, more accurate
data within a model not accounting for the presence of an accretion disc
in the system.

Based on the A5 spectral classification obtained by \citet{struve}, and 
the {\it spectral type -- effective temperature} calibration for Main 
Sequence stars \citep{harmanec}, we assumed 8150~K as the effective 
temperature of the primary star $T_{eff}^{prim}$. 
We also computed a second model for $T_{eff}^{prim}$=9500~K, as inferred 
by \citet{Olson93} from the colour change at the primary minimum. 
The spectroscopic mass ratio, $q_{spec} = 0.16$ (Section~\ref{bf}), 
was assumed, as a constant parameter. 
Theoretical values of albedo and limb darkening coefficients were 
adopted: $A_1=1$, $g_1=1$ for the radiative primary \citep{ziepel}, 
and $A_2=0.5$, $g_2=0.32$ for the convective secondary star \citep{lucy,ruc69}.  
We used the square root limb-darkening law, with respective coefficients 
from the \citet{claret} and \citet{diaz-cordoves} tables, 
applied as a function of the wavelength and effective temperature of the star. 
A simple analytical model of the reflection effect ($mref=1, nref=1$) 
was used during all computations.
We assumed the stars rotate synchronously.
We adjusted orbital inclination $i$, temperature of the secondary 
component $T_{eff}^{sec}$, potentials $\Omega_{1,2}$ and luminosity
$L_{1}$ (in the WD code notation) of the primary component.
The {\it IPB} control parameter was set to 0, therefore the luminosity 
of the secondary ($L_2$) was calculated as a function of temperatures, 
$L_{1}$, and the system geometry.

The solutions converged for both models, the one with higher (9500~K) temperature
of the primary component gave a somewhat better fit. The parameters 
obtained within the Roche model are given in Table~\ref{Tab.3}, while
the theoretical light curves along with observed ones are shown in 
Fig.~\ref{Fig.5} ( corresponding to the model with $T_{eff}^{prim}=8150$~K, 
see Sec.~\ref{discus} for explanation).
The theoretical light curves in the $V$ and $y$ filters fit the data best -- in 
other filters discrepancy with observations is larger, the secondary minimum 
being too shallow in longer wavelengths ($R$ and $I$), while too deep in $v$ and $b$ 
filters. 
The shape of the primary minimum is too shallow both in short and in long 
wavelengths. The ellipticity effect of the theoretical light curves roughly 
corresponds to that observed but, to achieve this, the resulting system configuration 
is detached with the primary star being well within its Roche lobe and 
the secondary component almost filling in its Roche lobe  ($\Omega_2=$2.2330 while 
$\Omega_{Roche}$=2.12996).
The solution could be further improved only by adding more free parameters; 
in order to obtain a better fit in the primary minimum, \citet{Olson93} 
considered over synchronous rotation of the primary star.

\subsection{An accretion disc model}
\label{thckdisk}

\begin{figure*}
\includegraphics[width=125mm,angle=-90]{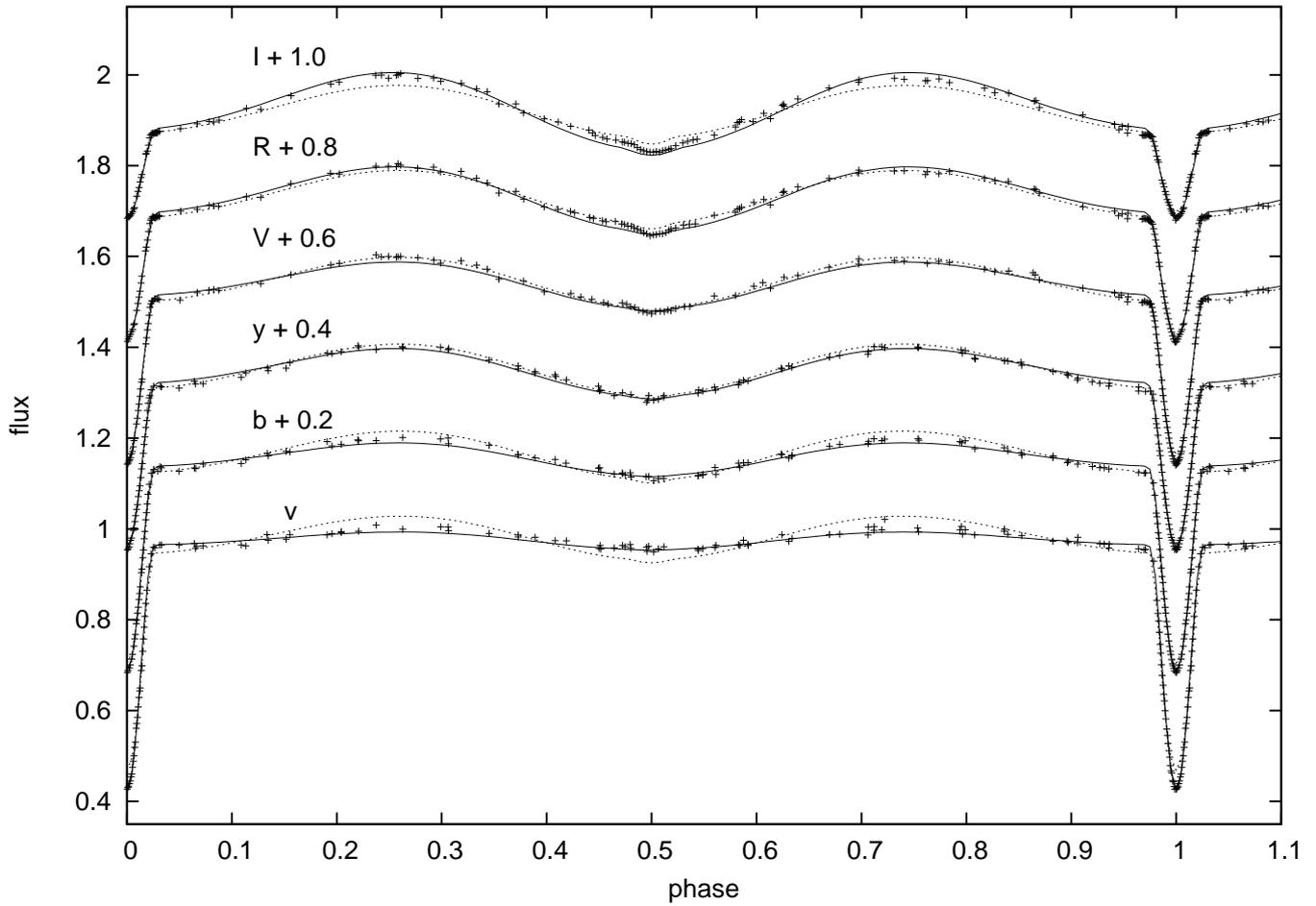}
\caption{The comparison between the observed (+) and synthetic light 
curves obtained from the stellar (dotted lines) and the disc models 
(continuous lines) assuming the primary temperature $T_{eff} = 8150$~K.}
\label{Fig.5}
\end{figure*}

As the next step, our goal was to derive a model consistent with all properties 
of \astrobj{WW And}, inferred both from the photometric and the spectroscopic data. 
Therefore, we applied the modified WD code which accounts for effects 
caused by a circular accretion disc surrounding the primary
component (\citealt{zola91}, and references therein). 
The disc thickness grows linearly with radius, it is assumed to be optically 
thick and radiate as a black body.
This model has already been successfully applied to solve the light curves
of some long-period Algol-type and W~Serpentis-type binary systems 
(\citealt{zola01}, and references therein). 
As mentioned in Section~\ref{intro}, the first, moderately successful
attempt to fit the {\it vbyI} light curves of \astrobj{WW And} using this code 
was published by \citet{zola97b}. The solution reasonably 
described the observed light curves for the photometric mass ratio $q=0.20$, but 
computations resulted in a very small disc radius r$_{d}$=0.06 (in units of component
separation). 

Similarly to the stellar model, we performed computations for two effective 
temperatures of the primary, mass accreting star: 9500~K and 8150~K. 
We kept the mass ratio fixed at its spectroscopic 
value (see Sec.~\ref{bf}) and a semidetached configuration was assumed, 
thus naturally providing matter for the accretion disc.
Furthermore, the disc radius was not adjusted but its value was chosen  
according to the orbital inclination as derived from the separation of blue 
and red peaks (Tab.~\ref{TabDiskRad}).
The orbital inclination $i$, potential of 
the primary component $\Omega_1$, disc luminosity $l_d$ and temperature $T_d$, 
$\beta$ -- the parameter describing the vertical disc thickness 
$z(r)=r \tan \beta$, luminosity of the primary $L_1$ and the effective 
temperature of the secondary star were adjusted in this step.

Models obtained for the two assumed temperatures of the primary star both
gave rather poor fits (as indicated by $\chi^2_{red,w}$).
The theoretical light curves showed too high ellipsoidal effects and an unacceptable, 
too deep secondary minimum, poorly resembling the data in all filters. 
The models also resulted in a cool secondary star ($T_{eff}^{sec}=3800$~K) and 
a negligible disc flux contribution in all filters except $V$ and $R$, 
where it reached 5\% and 13\% of the total system flux, respectively. 

This first approach gave us some hints about how to procede further. 
The assumed semidetached configuration caused higher 
than observed ellipsoidal effects, 
therefore the search procedure kept choosing a lower and lower temperature
of the secondary star to minimize these effects and eventually hit the lower
search limit. 
As argued by \citet{Olson93}, the secondary star cannot have such a low
temperature and, as a consequence, the temperature of the
mass-losing star was fixed at $T_{eff}^{sec}=4500$~K, estimated from
colour variations during the primary eclipse by \citet{Olson93}. 
Examination of the shape of the theoretical light curves around the secondary
eclipse (where the disc obscures the mass losing star) revealed that
if the disc was optically thick, its size was too big. 
In subsequent computations we added the disc radius to the list of adjusted 
parameters but kept the secondary star temperature fixed, thus the number 
of free parameters remained the same as in the previous step.
  
\begin{figure}
\includegraphics[width=63mm,angle=-90]{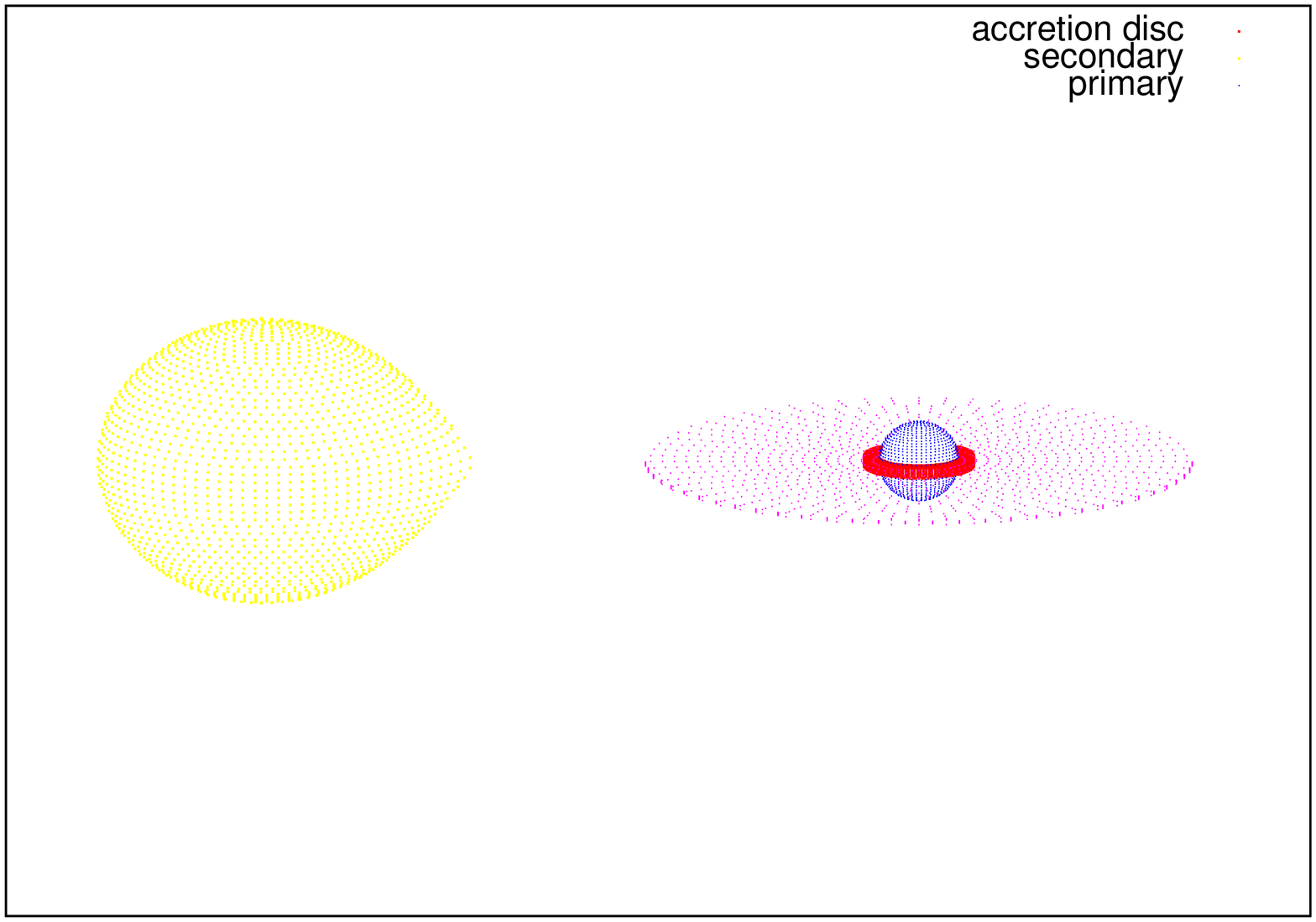}
\caption{The view of the model of \astrobj{WW And} as seen at the phase=0.25. The secondary
fills in its Roche lobe while the primary component is surrounded by an accretion
disc. The inner part of the disc (a ring close to the primary), up to the 
radius of 0.084 is optically thick while the outer region is optically thin. 
We made no attepmt to reproduce the vertical thickness of the optically thin 
region, it is shown only for displaying the entire disc size as indicated 
by the double peaked emission lines.}
\label{Fig.6}
\end{figure}

With the new setup, both sets of computations resulted in much better solutions.
Due to higher disc light contributions, the theoretical light curves fit the primary 
minimum very well in all filters, also the ellipsoidal effect agrees with 
that observed (see Fig.~\ref{Fig.5}). 
However, the solutions, independent of the assumed primary star temperature, 
converged at a small disc radius value -- 0.08-0.09 (in separation units) -- which 
agrees with the result of \citet{zola97b} 
but contradicts the disc size obtained from the H$_{\alpha}$ 
line red and blue peak separation measurements (Sec.~\ref{hmod}). 
Based on the above results we can draw the conclusion that, in order to fulfill 
the observed properties, the accretion disc in \astrobj{WW And} must be partly optically 
thin, with only the inner parts being optically thick (Fig.~\ref{Fig.6}).
Since our model assumes only an optically thick accretion discs,   
it cannot fully account for this case. 
Another problem encountered by the thick disc model is the very high disc 
light contribution in the $v$-filter -- more than 75\% of the total system
flux.

\section{Discussion and Conclusions}
\label{discus}

We gathered new $BVRI$ light curves of the Algol-type binary system 
\astrobj{WW And}, covering the entire orbital period and revealing 
for the first time the very well-defined shape of a grazing secondary minimum.
The Broadening Function method applied to the archival spectra allowed detection 
of the radial velocities of the primary star and estimate the system mass ratio
as $q_{spec}=0.16\pm0.03$.
        
The new light curves in the $VRI$ wide band filters, along with the Str{\"o}mgren
$vby$ filters published several years ago, have been modelled assuming both a stellar-only
model and another one accounting for the presence of an accretion disc
in the system. The solution derived with the stellar model confirmed the 
results published by \citet{Olson93} -- in order to obtain an acceptable fit, the
system configuration must be detached. This excludes the possibility of mass
transfer through the L$_{1}$ point as the source of matter for the disc manifesting
itself by the double-peaked hydrogen lines visible at all phases.
Our second approach, the application of a model accounting for disc effects (partial
obscuration of the mass-gaining star, eclipsing of the mass-losing star by the disc,
and the disc light contribution) and a semidetached configuration also produced
a reasonable fit to the observed light curves. 

Unexpected disc properties have
been deduced. In order to fit the photometric and spectroscopic data, only the
inner parts of the disc could be optically thick. The outer parts, from 
the radius of about 0.08  to the disc edge, must be optically thin to fit the shape
of the light curve near the secondary minimum. The disc light contribution needed to
reflect the observed ellipsoidal effects is high, especially in the Str{\"o}mgren 
$v$ filter where it reaches more than 75 percent. Such a high value, if it is not a 
numerical artifact -- it is very hard (if at all possible) to disentangle the
primary star light from that of the disc -- is not justified when one compares the
emitting surfaces of the disc and the stellar components. However, an additional
contribution could be possible due to an optically thin outer disc region: emission
in the Balmer lines and in the boundary layer visible at such a low inclination 
and relatively low disc thickness.  

We performed computations for two temperatures of the primary star: 8150~K 
and 9500~K. However, despite obtaining a slightly better fit for the higher 
T$_{eff}^{prim}$ value with the stellar model (with the disc model both 
temperatures provided equally good fits), we consider the lower  
temperature to be more probable and present solutions for 
T$_{eff}^{prim}$=8150~K (Table~\ref{Tab.3}).
This is for two reasons: (1) this value follows from the spectral classification 
and (more importantly), 
(2) only for this temperature can the spectrum of the primary star contain strong 
absorption lines in the Mg~II 4481\AA~region, allowing for determination 
of its radial velocities (described in Sec.~\ref{bf}).

We conclude that the disc model better explains the observed properties of
\astrobj{WW And}. Our results indicate that it consists of two stars having the following 
parameters derived from the combined spectroscopic and photometric data: 
the mass-gaining component, belonging to the Main Sequence ($R_1\approx3.25 R_{\odot}$, $M_1=3.18\pm0.09 M_{\odot}$),
and the mass-losing star -- separated by $53.0\pm0.5 R_{\odot}$ -- being 
a giant ($R_2\approx12.47 R_{\odot}$, $M_2=0.51\pm0.09 M_{\odot}$). 
The cooler star fills its Roche lobe, providing matter through the L$_{1}$ 
point at rather low rate to the disc surrounding the smaller, hotter star.
As the result, the disc has no stationary structure; instead, its outer parts 
are optically thin while the innermost region is optically thick. 
These inner, fast rotating parts of the disc could broaden the profile of 
the primary component seen in the Broadening Functions and in the spectra.
          
Further progress on understanding this system can be obtained from detailed
studies of the H$_{\alpha}$ and other Balmer emission lines, and from 
high-resolution spectra analyzed by means of the BF method. 
This would provide more precise 
radial velocities of the primary component, necessary to improve 
the preliminary value of the spectroscopic mass ratio, and allow for 
detailed studies of the accretion disc structure. 
The latter task will require modification of the disc model, as in 
its present form it can only be applied to a fully optically thick and circular 
disc, while an eccentric accretion disc could explain 
the asymmetries visible in the $VRI$ light curves around the secondary 
minimum and between the first and the second quadratures.

\section*{Acknowledgments}

We gratefully acknowledge Prof. Ed Olson and Dr. Paul Etzel for kindly provided spectra 
of \astrobj{WW And}, the anonymous referee for the very important comments and 
suggestions, and Dr. Greg Stachowski for language corrections. 
\newline
MS acknowledges the Canadian Space Agency Post-Doctoral position grant 
to Prof. Slavek M. Rucinski within the framework of the Space Science 
Enhancement Program.

\bibliographystyle{elsarticle-harv}



\end{document}